\documentclass[pre,preprint,onecolumn,showpacs,superscriptaddress,longbibliography]{revtex4-1}

\usepackage{graphicx}
\usepackage{dcolumn}
\usepackage{bm}
\usepackage{amssymb,amsfonts,amsmath}
\usepackage{color}
\usepackage{longtable}

\begin{document}

\title{Effect of Prandtl number on heat transport enhancement in Rayleigh-B\'enard convection under geometrical confinement}

\author{Kai Leong Chong}
\affiliation{Department of Physics, The Chinese University of Hong Kong, Shatin, Hong Kong, China}
\author{Sebastian Wagner}
\affiliation{Max Planck Institute for Dynamics and Self-Organization, Am Fassberg 17, 37077, Gottingen, Germany}
\author{Matthias Kaczorowski}
\affiliation{Department of Physics, The Chinese University of Hong Kong, Shatin, Hong Kong, China}
\author{Olga Shishkina}
\affiliation{Max Planck Institute for Dynamics and Self-Organization, Am Fassberg 17, 37077, Gottingen, Germany}
\author{Ke-Qing Xia}
\affiliation{Department of Physics, The Chinese University of Hong Kong, Shatin, Hong Kong, China}

\date{\today}

\begin{abstract}
We study, using direct numerical simulations, the effect of geometrical confinement on heat transport and flow structure in Rayleigh-B\'enard convection in fluids with different Prandtl numbers. Our simulations span over two decades of Prandtl number $Pr$, $0.1 \leq Pr \leq 40$, with the Rayleigh number $Ra$ fixed at $10^8$. The width-to-height aspect ratio $\Gamma$ spans between $0.025$ and $0.25$ while the length-to-height aspect ratio is fixed at one. We first find that for $Pr \geq 0.5$, geometrical confinement can lead to a significant enhancement in heat transport as characterized by the Nusselt number $Nu$. For those cases, $Nu$ is maximal at a certain $\Gamma = \Gamma_{opt}$. It is found that $\Gamma_{opt}$ exhibits a power-law relation with $Pr$ as $\Gamma_{opt}=0.11Pr^{-0.06}$, and the maximal relative enhancement generally increases with $Pr$ over the explored parameter range. As opposed to the situation of $Pr \geq 0.5$, confinement-induced enhancement in $Nu$ is not realized for smaller values of $Pr$, such as $0.1$ and $0.2$. The $Pr$ dependence of the heat transport enhancement can be understood in its relation to the coverage area of the thermal plumes over the thermal boundary layer (BL) where larger coverage is observed for larger $Pr$ due to a smaller thermal diffusivity. We further show that $\Gamma_{opt}$ is closely related to the crossing of thermal and momentum BLs,  and find that $Nu$ declines sharply when the thickness ratio of the thermal and momentum BLs exceeds a certain value of about one. In addition, through examining the temporally averaged flow fields and 2D mode decomposition, it is found that for smaller $Pr$ the large-scale circulation is robust against the geometrical confinement of the convection cell.
\end{abstract}

\pacs{47.27.T-,44.25.+f,47.27.De}

\maketitle

\section{Introduction} \label{intro}
Thermally-driven flows are ubiquitous phenomena in nature and industrial applications. Turbulent Rayleigh-B\'enard (RB) convection in a fluid layer heated from below and cooled from above is the classical model for studying such phenomena. 
This model has been used to investigate important issues related to the heat transport and flow dynamics in a broad range of research fields, including astrophysics, geophysics and engineering \citep{busse1970aj,wyngaard1992atmospheric,linden1999arfm,shishkina2012jt, Bailon2012}. Over the past decades, extensive studies of RB convection have been conducted using experimental, numerical and theoretical approaches  ~\citep{ahlers2009rmp,lohse2010arfm,chilla2012epj,xia2013taml}. In RB convection, the control parameters that govern the flow are the Rayleigh number $Ra$ and the Prandtl number $Pr$. Besides, the geometry of the container also plays an important role, in particular, the diameter-to-height aspect ratio $\Gamma$. The studies of RB convection usually concern about the global heat transport across the system and also the problem of heat flow optimization. It is particularly important in passive thermal management, which is sometimes indispensable in industrial and engineering applications. Various methods to passively enhance heat transport have been found in RB studies. For instance, for RB cells with rough surfaces \citep{shen1996prl,du1998prl,wei2014jfm,Wagner2015,xie2017jfm} that the heat transport can be enhanced significantly as the roughness modifies the thermal boundary layers, leading to more frequent emission of the thermal plumes. Examples also include RB flows with polymer additives \citep{ahlers2010prl,benzi2010prl,wei2012pre,benzi2012jfm,xie2015jfm}. Furthermore, the heat transport can also be enhanced by adding a stabilizing force in addition to thermal driving such that highly coherent thermal plumes are formed \citep{huang2013prl,Horn2014,chong2015prl,chong2017prl}.

Large number of RB studies in the past decade \citep{grossmann2003jfm,ching2006jt,bailon2010jfm,poel2012pf,zhou2012jfm,huang2013prl,wagner2013pof,chong2015prl,chong2016jfm,chong2017prl} have been devoted to the investigation of how the varying geometrical control parameter influence the heat transport and flow properties. They can be separated into two categories: One with the aspect ratio larger than one and the other with the aspect ratio smaller than one. The present paper focuses on the latter situation, i.e. RB under geometrical confinement. Huang et al. \cite{huang2013prl} has found that for RB cell with $Pr=4.38$  at constant $Ra$, the Nusselt number $Nu$ can increase on decreasing the width-to-height aspect ratio $\Gamma$, while the flow strength is reduced monotonically at the same time. The numerical work by Chong et al. \cite{chong2015prl} has further studied a broader range of parameters with $1/64 \leq \Gamma \leq 1$, $10^7 \leq Ra \leq 10^{10}$ at $Pr=4.38$. They have revealed that the confinement-induced heat transport enhancement only occurs over a particular range of $\Gamma$ for a given $Ra$. For weakly confined geometry, $Nu$ is found to be insensitive to the decrease in $\Gamma$ until the cell width becomes smaller than the average spacing between the thermal plumes near the thermal boundary layers. In other words, there exists an onset aspect ratio for enhancement which is given by $\Gamma_c=12.42Ra^{-0.21}$ \cite{chong2015prl}. 
When $\Gamma$ is below the onset value, the so-called plume-controlled regime sets in. Within this regime thermal plumes condensate into giant or super plumes at the opposite boundary layers, which enables them to more efficiently cool down or heat up the corresponding plate. In this regime, $Nu$ increases continuously on decreasing $\Gamma$ until the severely-confined regime is entered \cite{chong2016jfm}. The boundary between the two regimes suggest an optimal aspect ratio at which $Nu$ is maximized and the dependency of $\Gamma_{opt}$ on $Ra$ is given by $\Gamma_{opt}=29.37Ra^{-0.31}$. On the other hand, the study by Wagner \& Shishkina \cite{wagner2013pof} revealed that the existence of the plume-controlled regime depends strongly on $Pr$. For example, for $Pr=0.786$ with $1/10 \leq \Gamma \leq 1$ and $10^5 \leq Ra \leq 10^7$, no significant enhancement in $Nu$ was found. All this calls for an in-depth study on the influence of $Pr$ on confinement-induced $Nu$ enhancement. Another issue to be studied is the flow dynamics, which depends strongly on the geometrical control parameters. For a convection cell of the aspect ratio around one, it is known that there exists a persistent large-scale circulation (LSC) of a single-roll flow pattern \citep{cioni1997jfm,brown2005prl,sun2005prl,xi2006pre}. However, LSC becomes horizontally adjacent multiple rolls for the aspect ratio much larger than one \citep{funfschilling2005jfm,sun2005jfm,bailon2010jfm, xia2008large}. Whereas LSC becomes unstable when the aspect ratio is reduced less than one as reflected by the increased flow reversals or cessations, for both 3D \cite{xi2008flow, xi2008azimuthal} and quasi-2D configurations \citep{ni2015jfm,huang2016jfm}.

In this paper, we present a comprehensive direct numerical simulation (DNS) study of the effect of $Pr$ on heat transfer enhancement for RB convection under geometrical confinement. In addition to heat transport, we also analyze how $Pr$ influences the  confinement-induced change in global flow structures. The rest of the paper is organized as follows. In Sec.\ref{numerical}, we describe the numerical method and the simulation parameters. In Sec.\ref{result}, we first present the dependency of global Nusselt number and Reynolds number on the aspect ratio for different $Pr$. Further, an analysis on the local quantities, such as temperature fluctuation, velocity fluctuation, boundary layers thickness are presented. Then, we compare the global flow structures for different $Pr$ and $\Gamma$ qualitatively by temporally averaged flow fields and quantitatively by 2D mode decomposition. In Sec.\ref{conclusions}, we summarize our findings.

\begin{figure}[!h]
\includegraphics[width=0.4\textwidth]{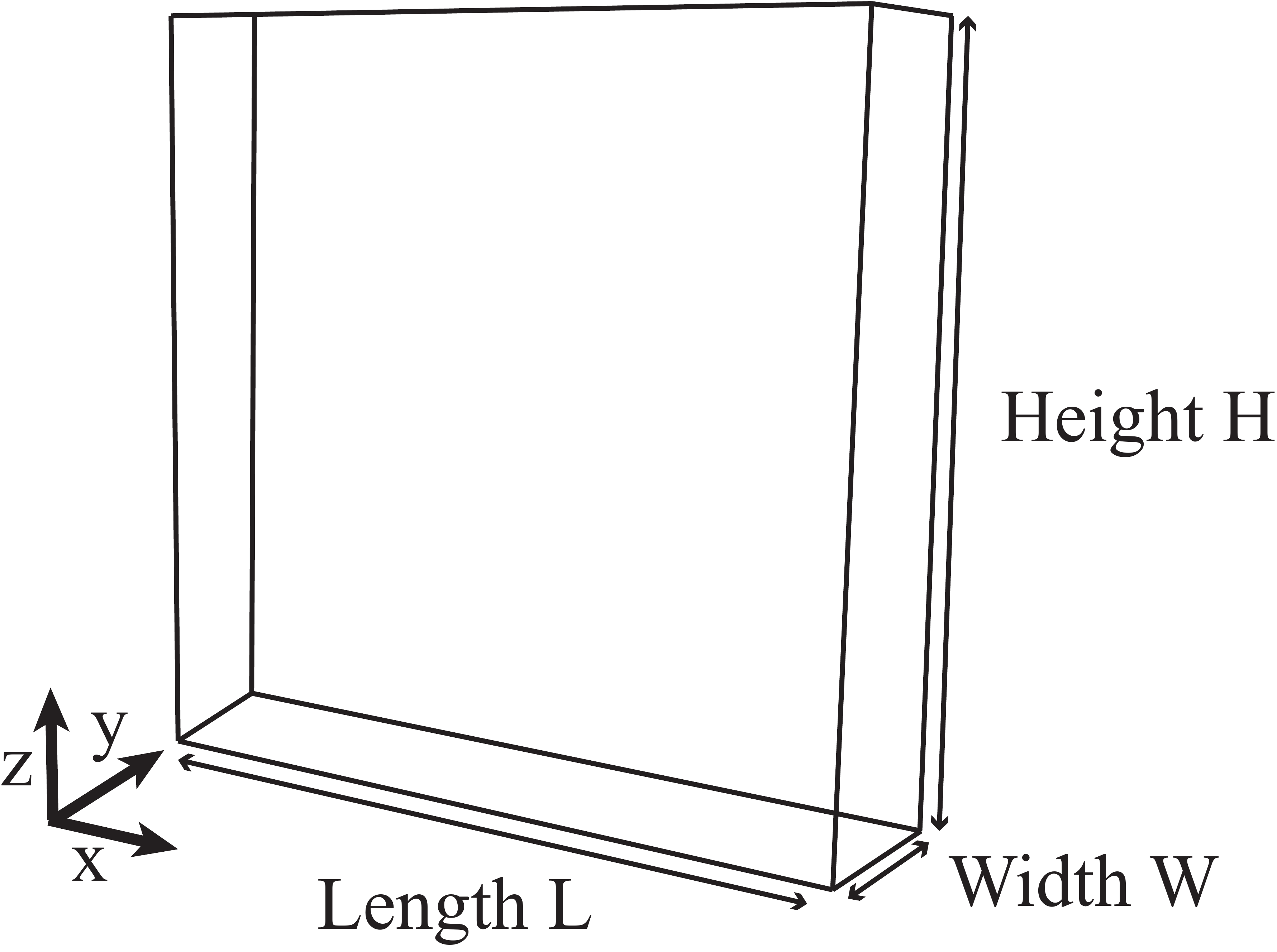}
\caption{\label{fig:sch}Schematic plot of the RB set-up.}
\end{figure}

\begin{figure}[!h]
\includegraphics[width=0.9\textwidth]{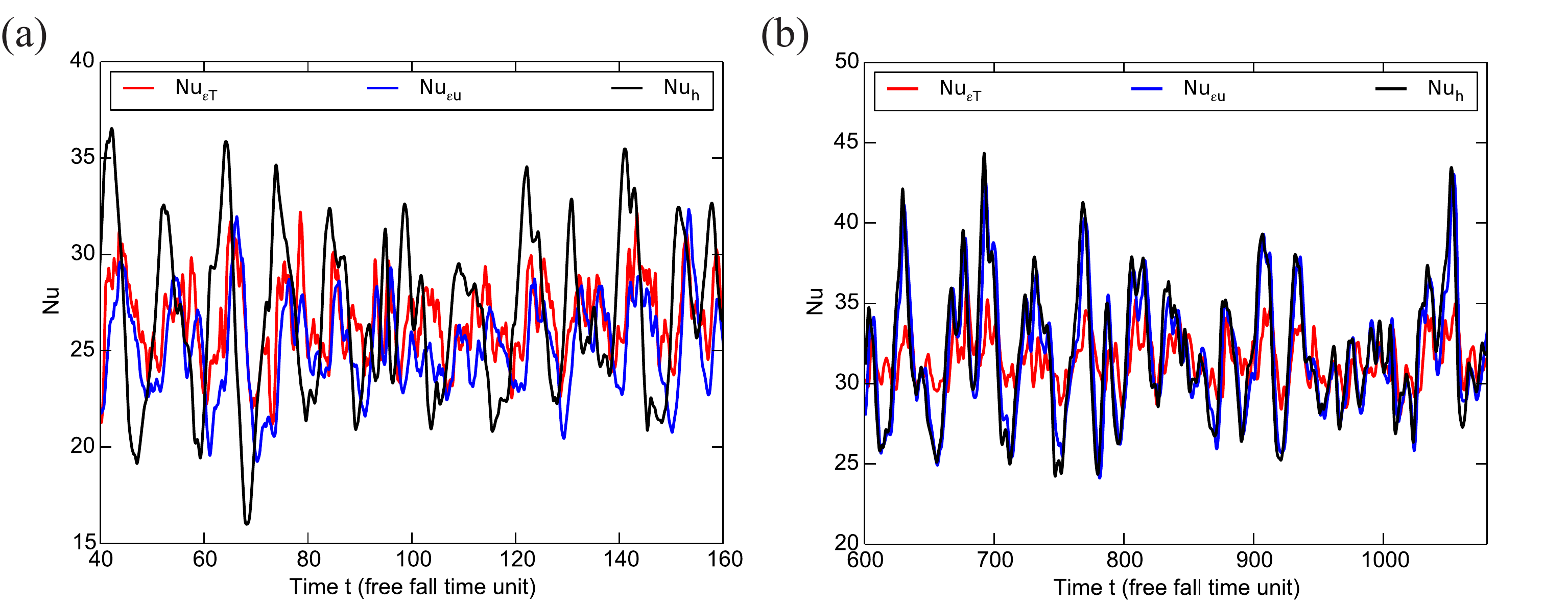}
\caption{\label{fig:timeseries}
Segment of Nusselt numbers versus time after reaching statistical steady state for (a) $Pr=0.1$, $\Gamma=0.25$ and $Ra=10^8$ and (b) $Pr=10$, $\Gamma=0.25$ and $Ra=10^8$. 
Three different curves show the time series of Nusselt number calculated by globally averaged viscous dissipation ($Nu_{\epsilon_u}$), thermal dissipation ($Nu_{\epsilon_T}$) and layer averaged of heat flux respectively ($Nu_{h}$), see the main text.}
\end{figure}

\section{Numerical methods and simulation parameters} \label{numerical}
The velocity field $\textbf{u}$ is described by the Navier--Stokes equation within the Oberbeck-Boussinesq approximation together with the incompressibility condition. The temperature field $T$ satisfies the advection-diffusion equation. The non-dimensional form of these equations is given by:
\begin{equation}
\partial\textbf{u}/\partial t + \textbf{u} \cdot \nabla \textbf{u} + \nabla p = {(\rm{Pr} /{\rm{Ra}})^{1/2}}{\nabla ^2}\textbf{u} + \rm{T} \textbf{z},
\end{equation}
\begin{equation}
\partial\rm{T}/\partial t + \textbf{u} \cdot \nabla \rm{T} = {(\rm{Pr}\rm{Ra})^{ - 1/2}}{\nabla ^2}{\rm{T}},
\end{equation}
\begin{equation}
\nabla \cdot \textbf{u} = 0,
\end{equation}
where the dimensionless control parameters are the Rayleigh number $Ra=\hat{\beta} \hat{g} \Delta \hat{T} \hat{H}^3/\hat{\nu}\hat{\kappa}$ and the Prandtl number $Pr=\hat{\nu}/\hat{\kappa}$. Here $\hat{\beta}$, $\hat{\nu}$, $\hat{\kappa}$ denote the thermal expansion coefficient, kinematic viscosity and thermal diffusivity of the fluid and $\hat{g}$ is the gravitational acceleration acting vertically. Also, $\Delta \hat{T}=\hat{T}_{bottom}-\hat{T}_{top}$ denotes the temperature difference between top and bottom plate separated by cell height $\hat{H}$. The physical quantities are sought in dimensionless form with the cell height $\hat{H}$ for length scale, and free-fall time $(\hat{H}/(\hat{\beta} \hat{g} \Delta \hat{T}))^{1/2}$ for time scale, and the velocities are normalized by the free-fall velocity $(\hat{\beta} \hat{g} \Delta \hat{T} \hat{H})^{1/2}$. The temperature are made dimensionless by $T=(\hat{T}-\hat{T}_m)/\Delta \hat{T}$ with $\hat{T}_m=(\hat{T}_{bottom}+\hat{T}_{top})/2$ and $\hat{T}$ being the dimensional temperature. The DNS are conducted in a box, presented in Fig.~\ref{fig:sch} together with the nomenclature and coordinates. For the domain boundaries, all walls are set to be no-slip and impermeable. The vertical walls are adiabatic while the top and bottom plates are isothermal with $T_{top}=-0.5$ and $T_{bottom}=0.5$ after normalization. 

The equations are solved by a fourth-order finite-volume method on staggered grids. 
The G\"ottingen group used the {\sc Goldfish} code as in \cite{Wagner2015, Shishkina2016},
while the Hong Kong group used their well-tested extension \citep{kaczorowski2013jfm,kaczorowski2014jfm} of the code \citep{kaczorowski2008nrnefm6}. A requirement for obtaining reliable results in DNS studies is to resolve the Kolmogorov ($\eta_k$) and the Batchelor ($\eta_b$) length  scales. The global estimation of both scales in dimensionless form are given by $\eta_k=\sqrt{\rm{Pr}}/(\rm{Ra}(\rm{Nu}-1))^{1/4}$ and $\eta_b=1/(\rm{Ra}(\rm{Nu}-1))^{1/4}$. From these relations we can see that the smallest length scale to be resolved for $Pr>1$ is $\eta_b$, whereas for $Pr<1$, $\eta_k$ becomes the smallest scale. Also, sufficient resolution is needed inside the boundary layers as suggested by Shishkina {\it et al. }\citep{shishkina2010njp} and thus the non-uniform mesh with denser grid points in the boundary layer regions is adopted in our simulations. Based on the above requirements, $768$ vertical grid points were used for $Pr=0.1$ and $256$ vertical grid points for $Pr=40$. The statistical data are collected after statistical steady state has been reached as judged by the convergence of global $Nu$. Figure \ref{fig:timeseries} shows examples of $Nu$ time series after reaching statistical steady state from where the oscillation about certain mean value is seen.

We present the simulations with $90$ cases in total. To study the effect of $Pr$ on change of heat transfer and flow dynamics brought by geometrical confinement, over two decades of $Pr$ ($0.1 \leq Pr \leq 40$) have been covered and all fixed at the same $Ra$ which is $10^8$. The width-to-height aspect ratio $\Gamma=W/H$ has been varied from $0.025$ to $0.25$ while the length-to-height aspect ratio is fixed at $1$. The details of all the cases including their meshes and check of resolution requirements are summarized in table \ref{tab:sim}.

\section{Results and discussion} \label{result}
\subsection{Nusselt number and Reynolds number}
\begin{figure}[!h]
\includegraphics[width=0.9\textwidth]{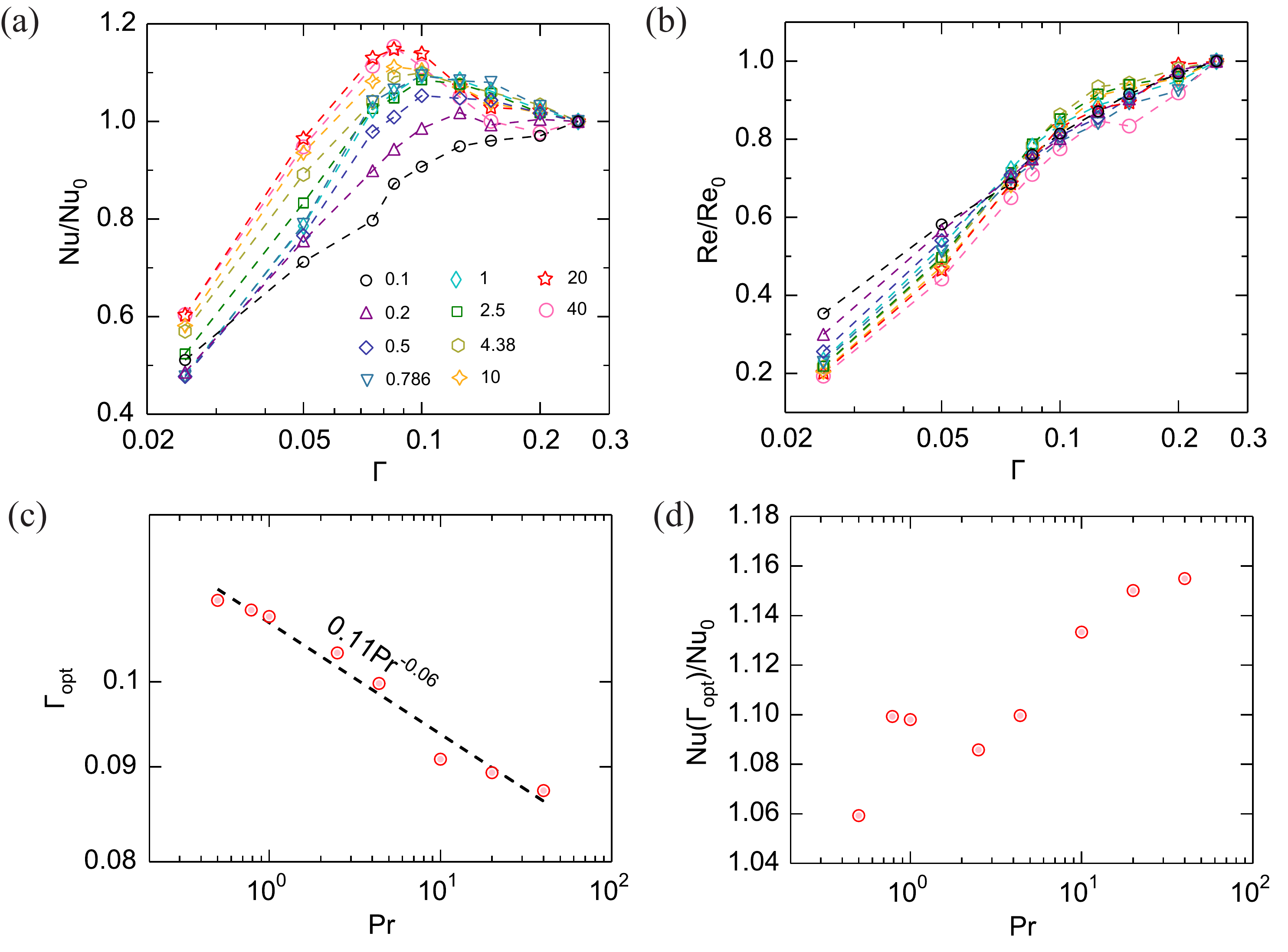}
\caption{\label{fig:NuReG}Normalized (a) global Nusselt number and (b) Reynolds number versus $\Gamma$, for different $Pr$, where $Nu_0$ and $Re_0$ are the values obtained at $\Gamma=0.25$; (c) the optimal aspect ratio $\Gamma_{opt}$ (if exists) versus $Pr$, where the dashed line represents the best power-law fit $\Gamma_{opt}=0.11Pr^{-0.06}$; (d) the maximum $Nu$ enhancement $Nu(\Gamma_{opt})/Nu_0$ versus $Pr$.}
\end{figure}

We first examine how Nusselt number $Nu$ and Reynolds number $Re$ vary with decreasing $\Gamma$ for fluid with different $Pr$. The evaluation of $Nu$ is based on three different methods. The first one is to estimate global $Nu$ through the formula $Nu_h=\langle(RaPr)^{1/2}u_z T - \partial T/ \partial z \rangle_{x,y,t}$ which represents the temporally averaged heat flux across a horizontal plane. By taking average of $Nu_h$ across every horizontal plane, we can obtain the first estimation of $Nu$. Other ways to estimate $Nu$ make use of the exact relations $Nu=\langle\epsilon_u\rangle(RaPr)^{1/2}+1$ and $Nu=\langle\epsilon_T\rangle(RaPr)^{1/2}$ where $\langle\epsilon_u\rangle$ and $\langle\epsilon_T\rangle$ represent, respectively, the viscous and thermal dissipation rates averaged over time and the entire volume \citep{shraiman1990pra,grossmann2000jfm}. We obtain the numerically measured $Nu$ by their mean value and the error of $Nu$ by their standard deviation. The evaluation of $Re$ is based on the formula $Re=\sqrt{\langle\textbf{u}^2\rangle(Ra/Pr)}$, where $\langle\textbf{u}^2\rangle$ represents the root-mean-square value of velocities averaged over time and entire domain.

The result of normalized $Nu$ versus $\Gamma$ for a given $Pr$ is shown in figure \ref{fig:NuReG} (a). It is seen that the system's response to confinement has a strong $Pr$ dependence that can be classified as follows. The cases with $Pr \geq 0.5$ belong to one group, for which a regime with significant enhancement of $Nu$ exists upon decreasing $\Gamma$. The cases with $Pr = 0.1$, $0.2$ belong to the second group for which no significant $Nu$ enhancement is seen. We also find that the Reynolds number $Re$ exhibits a large and monotonic decrease with $\Gamma$,  as shown in Fig.~\ref{fig:NuReG} (b). These results are in agreement with the previous finding that slower flow can indeed transport more heat \citep{huang2013prl}. For the cases with enhanced heat transport, $Nu$ reaches a maximum value upon further confinement, after which heat transport efficiency declines sharply and the optimal aspect ratio $\Gamma_{opt}$ can be defined through this trend. 

\begin{figure}[!h]
\includegraphics[width=0.8\textwidth]{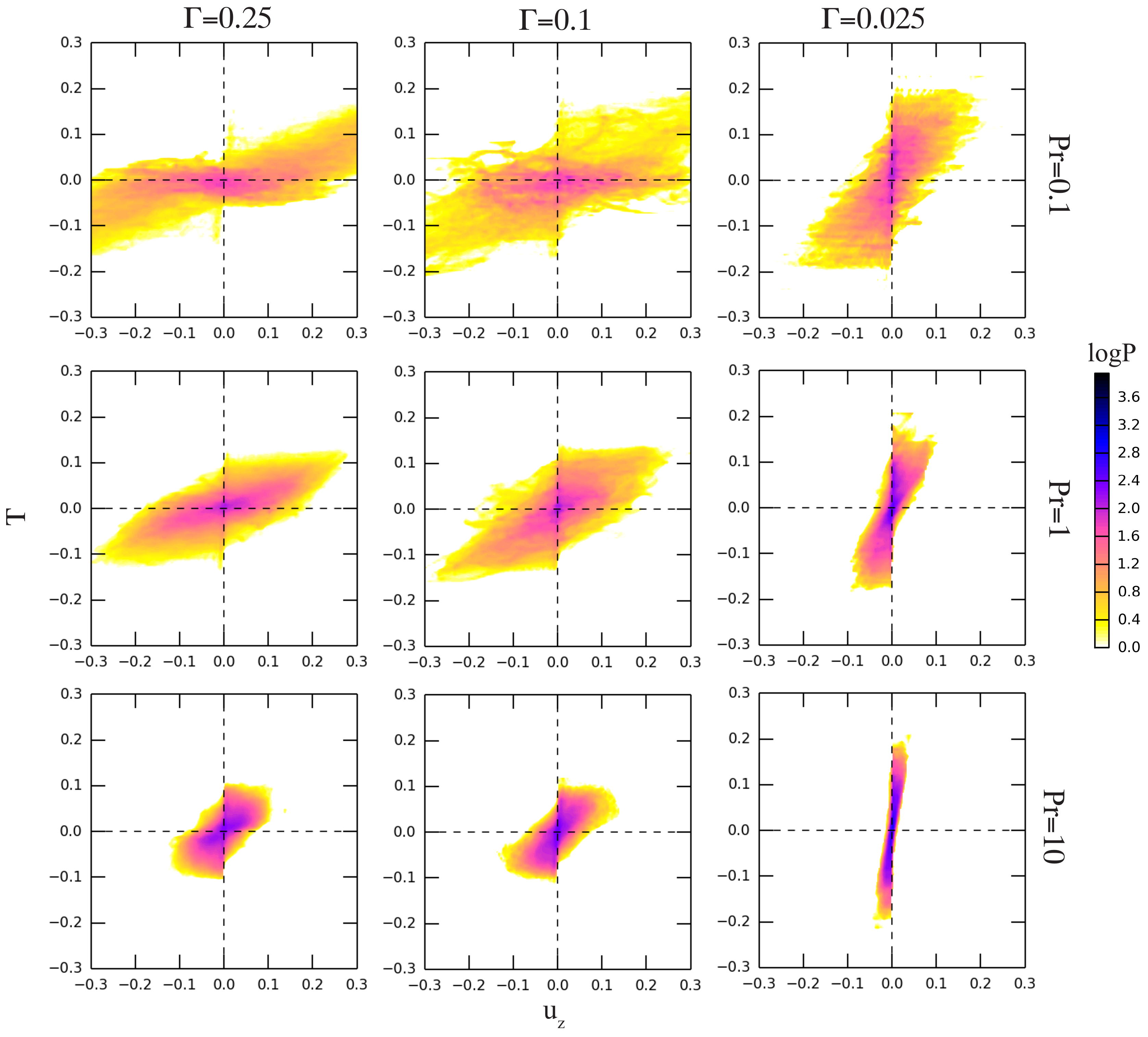}
\caption{\label{fig:jpdf}Joint probability density function p.d.f.s for the temperature $T$ and vertical velocity $u_z$ (both are dimensionless) evaluated at mid-height for $Pr=0.1$, $1$ and $10$ with $\Gamma=0.25$, $0.1$ and $0.025$.}
\end{figure}

In \cite{chong2015prl} by example of $Pr=4.3$ it was shown that the optimal aspect ratio $\Gamma_{opt}$, which provides the maximal $Nu$, exists for any $Ra$. The present results suggest that the optimal $\Gamma_{opt}$ also exists for any $Pr$. This fact is very important for optimization of heat transport in practical applications.

Figure \ref{fig:NuReG} (c) shows the dependence of $\Gamma_{opt}$ on $Pr$ in a log-log plot. In order to estimate $\Gamma_{opt}$ more accurately, quadratic fit has been made to the three points in the neighborhood of the peak position in figure \ref{fig:NuReG} (a). The figure reveals that there is a power-law relationship among the two quantities, and the fitting of respective data yields $\Gamma_{opt}=0.11Pr^{-0.06}$. It is worthwhile to recall that the relationship between $\Gamma_{opt}$ and $Ra$ has been found previously \citep{chong2015prl}, where $\Gamma_{opt}$ and $Ra$ follow a much stronger power-law relation: $\Gamma_{opt}=29.37Ra^{-0.31}$. In figure \ref{fig:NuReG} (d) we further examine the maximum enhancement versus $Pr$. The data is somewhat scattered, but a general trend is the growth with $Pr$ of the relative heat transport enhancement within the explored parameter range. Specifically, for the lowest value of $Pr$($=0.5$) the enhancement is $\sim 5.3\%$, while for the highest $Pr$($=40$) studied the enhancement is $15.3\%$.

\subsection{Joint probability density function of velocity and temperature fluctuations}

To gain further insights into how the bulk flow is modified by geometrical confinement, especially for fluid with different $Pr$, we now focus on the local quantities. One of such local measurements revealing bulk flow properties is the joint probability density function between the temperature (vertical axis) and vertical velocity (horizontal axis) at mid-height of the cell, which is shown in Fig.~\ref{fig:jpdf}. The figure can be interpreted from two perspectives; either at fixed $\Gamma$ with varying $Pr$ or  at fixed $Pr$ with varying $\Gamma$. First, along the column as $Pr$ increases, the shape of distribution function shrinks horizontally, meaning for larger $Pr$ the extreme events for velocity become less probable. This feature can be understood as the fluid flow becomes less turbulent for larger $Pr$.
Second, along the row with fixed $Pr$ but decreasing $\Gamma$, the shape change of distribution functions lead us to identify two competing effects. On one hand geometrical confinement slows down the bulk flow due to stronger drag from the sidewalls, and hence it is less probable for large velocity to occur. On the other hand, confinement leads to the formation of highly coherent plumes \citep{huang2013prl} and thus the temperature distribution is elongated since hot (cold) plumes become hotter (colder) when they reach the mid-height. However, the bulk properties observed here cannot explain why $Nu$ enhancement has not been realized for $Pr=0.1$ and $0.2$. As can be seen increasing plume coherency in the bulk has also been realized for $Pr=0.1$ but in such case, there is no heat transport enhancement globally. This observation suggests that instead of bulk property, other quantities may also play important roles on heat transport enhancement, for example the local properties at the edge of BL.

\subsection{Physical quantities in thermal boundary layers}
\begin{figure}[!h]
\includegraphics[width=0.9\textwidth]{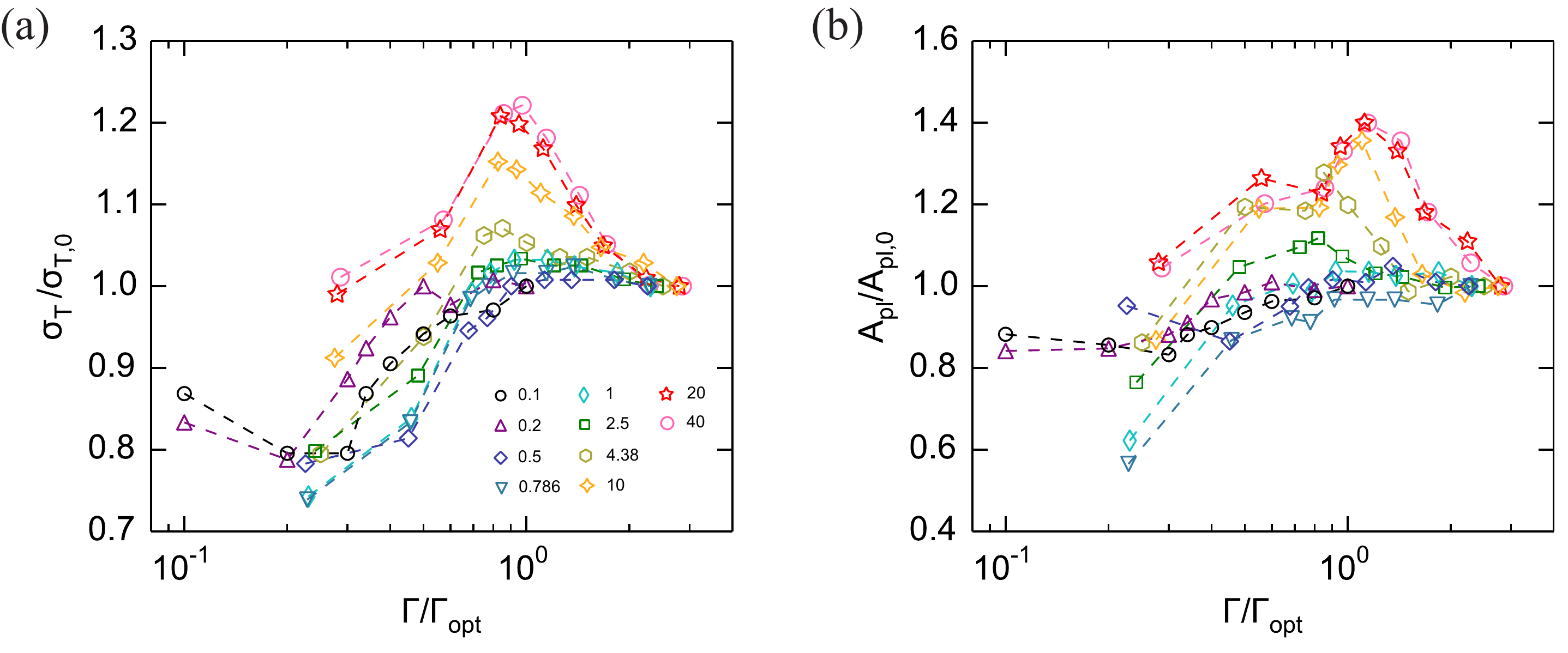}
\caption{\label{fig:plcov} (a) Temperature standard deviation $\sigma_T$ normalized by that obtained at $\Gamma=0.25$ denoted as $\sigma_{T,0}$ versus rescaled aspect ratio $\Gamma/\Gamma_{opt}$ evaluated at the edge of thermal BL. Note that $\Gamma_{opt}$ is ill-defined for $\Gamma=0.1$ and $0.2$, and we take $\Gamma_{opt}=0.25$ for convenience. (b) Areal coverage of cold plumes over the edge of the (hot) bottom thermal BL $A_{pl}$ normalized by its value obtained at $\Gamma=0.25$ (denoted as $A_{pl,0}$) versus $\Gamma/\Gamma_{opt}$.}
\end{figure}

\begin{figure}[!h]
\includegraphics[width=0.9\textwidth]{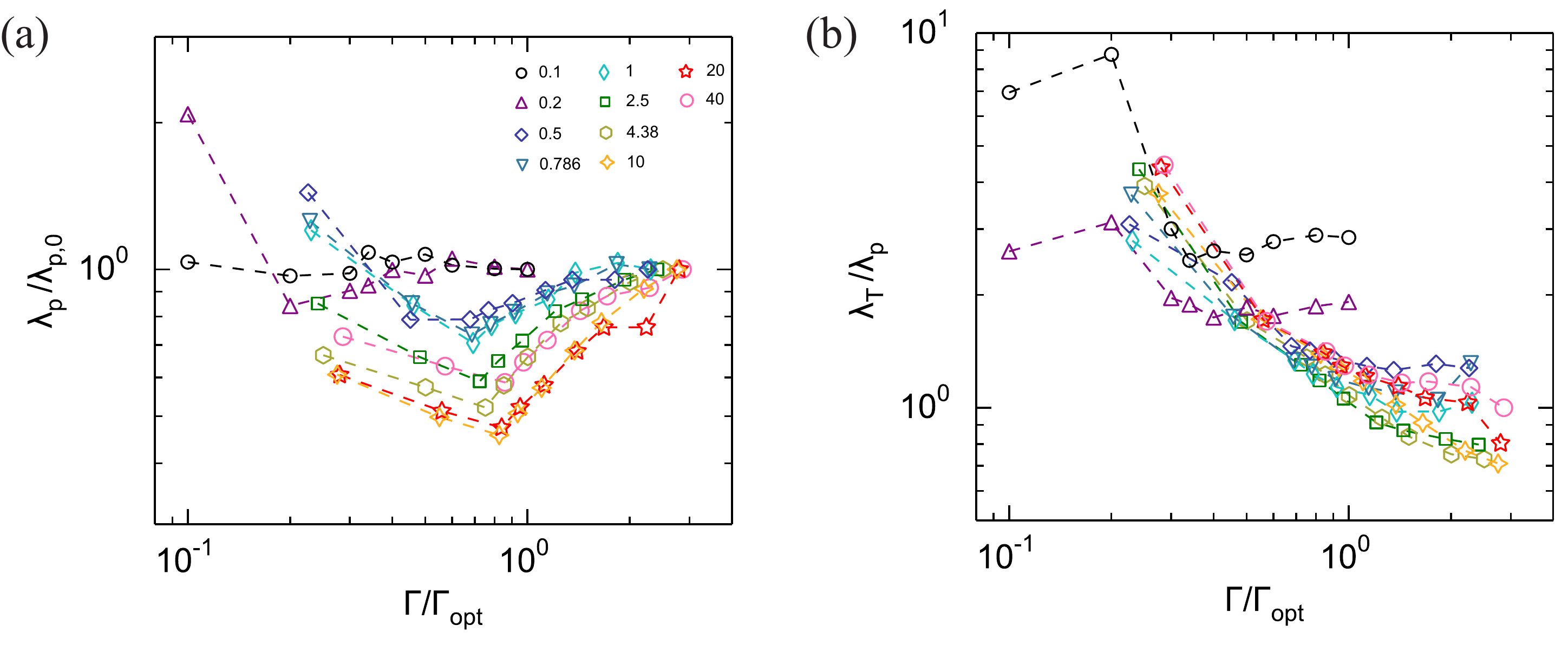}
\caption{\label{fig:bdcrossing} (a) Momentum boundary layer thickness $\lambda_p$ normalized by that obtained at $\Gamma=0.25$ (denoted as $\lambda_{p,0}$) versus rescaled aspect ratio $\Gamma/\Gamma_{opt}$. (b) The ratio of temperature boundary layer thickness to momentum one $\lambda_T/\lambda_p$ versus $\Gamma/\Gamma_{opt}$.}
\end{figure}

Figure \ref{fig:plcov} (a) shows temperature standard deviation $\sigma_T$ evaluated at the edge of thermal BL versus $\Gamma$/$\Gamma_{opt}$. We normalize $\Gamma$ in this way to better illustrate how the quantities concerned vary as the optimal point is approached. Here we estimate the thickness of the temperature BL based on the vertical temperature standard deviation profiles from which the location of maximum value is regarded as the thickness. The local quantities taken from the thermal BL indeed allow us to observe the differences caused by $Pr$. For large $Pr$ cases, we clearly see that $\sigma_T$ increases as $\Gamma$ is reduced towards $\Gamma_{opt}$. Also worthy of noting is that the trend is more pronounced for the few largest values of $Pr$. It may be understood by recognizing that larger $Pr$ corresponds to smaller thermal diffusivity and so the plumes are able to better preserve their heat content when traversing to the opposite end. In case of too large thermal diffusivity, i.e. cases of $Pr=0.1$ and $0.2$, the plume's heat content loss to ambient fluid is most severe. Indeed, we do not observe appreciable increase of $\sigma_T$ at thermal BL for those cases.

Besides temperature standard deviation at the edge of thermal BL, plume coverage is also important for heat transport enhancement by confinement as shown in \citep{chong2015prl}. The estimation of plume coverage requires the extraction of cold plumes over the hot bottom BL and is defined as the area satisfying $-(T-\langle T \rangle_{x,y}) \geq cT_{rms}$, and the rms is the value for $\Gamma=0.25$  at a given $Pr$. The empirical parameter $c$ is chosen to be 0.5. We have tested different choices of $c$ and found that  our main conclusions do not depend on the particular choice of its value. Figure \ref{fig:plcov} (b) shows the normalized coverage of cold plumes over the edge of the (hot) bottom thermal BL versus rescaled aspect ratio $\Gamma/\Gamma_{opt}$. For large $Pr$ cases, it shows clearly that, as $\Gamma_{opt}$ is approached, the plume coverage is increased significantly. This suggests that the bottom (top) BL is cooled down (heat up)  more efficiently, which also results in a thinner BL. Again, such increase in plume coverage is largely absent for very low $Pr$, i.e., $Pr=0.1$ and $0.2$. Thus the properties of both $\sigma_T$ and plume coverage at the edge of the thermal BL can explain the behavior of $Pr$ dependence  of $Nu$ in response to confinement.

\subsection{Crossing of momentum and temperature boundary layers}
\begin{figure}[!h]
\includegraphics[width=0.9\textwidth]{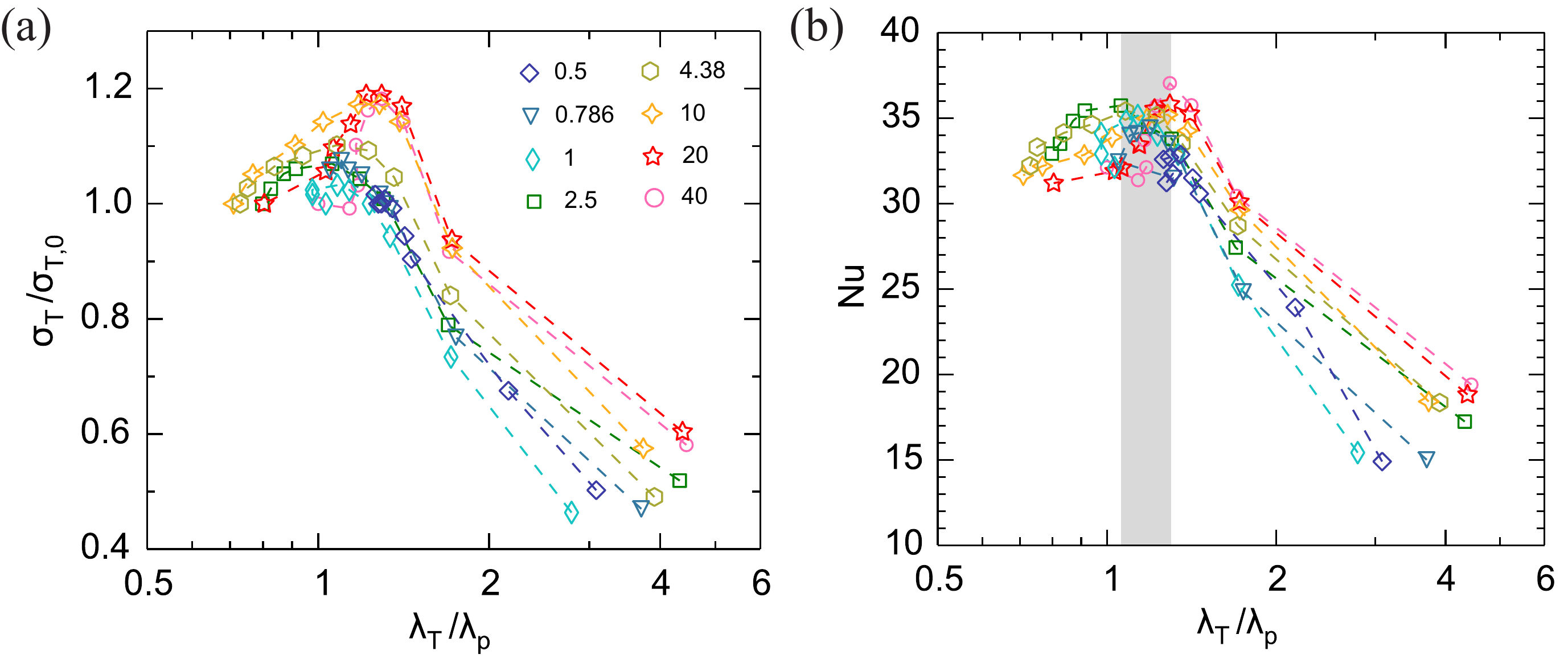}
\caption{\label{fig:bdcrossingNu}(a) Normalized temperature standard deviation $\sigma_T/\sigma_{T,0}$ versus thickness ratio $\lambda_T/\lambda_p$, where $\sigma_T$ is evaluated at the edge of momentum BL. (b) Nusselt number $Nu$ versus $\lambda_T/\lambda_p$.}
\end{figure}
\begin{figure}[!h]
\includegraphics[width=0.8\textwidth]{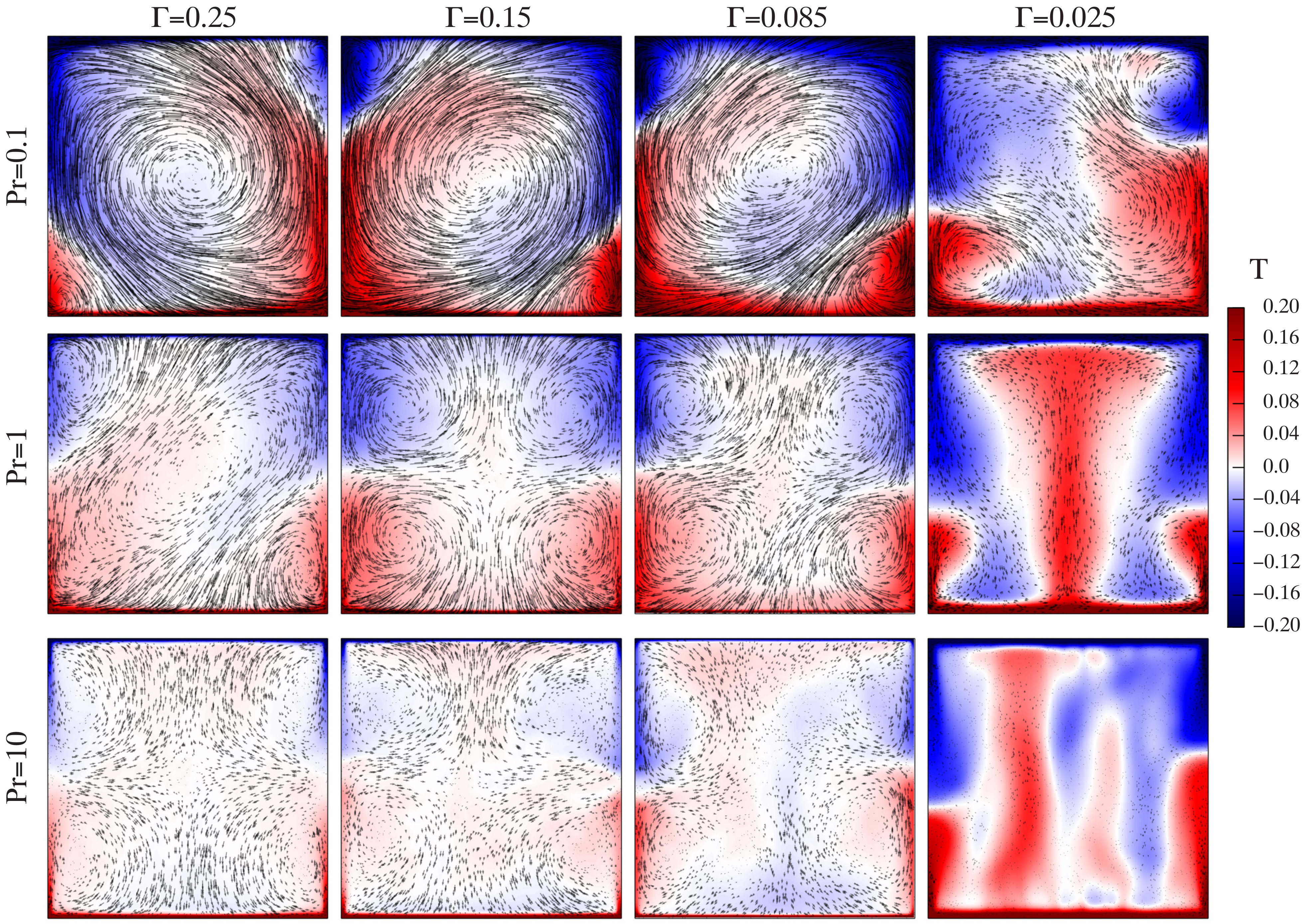}
\caption{\label{fig:meanfield}Time-averaged temperature and velocity fields at mid-way along the confinement direction for three different Pr, which are $Pr=0.1$, $1$ and $10$, and for four different $\Gamma$, which are $\Gamma=0.25$, $0.15$, $0.085$ and $0.025$. The magnitude of the velocity is represented by the length of the arrows in non-dimensional unit and the temperature is coded in color.}
\end{figure}

In a recent study it has been shown that the relative thickness of thermal and momentum BLs  plays  a major role on optimal transport \citep{chong2017prl}. 
To understand the $Pr$-dependent optimal point found in this work, we follow the same idea. Using the proposal made by Chong {\it et al.} the thickness of momentum BL $\lambda_p$ may be defined by the first peak of $(\partial_x u)^2+(\partial_y v)^2+(\partial_z w)^2$ profile  \citep{chong2017prl}. The edge of momentum BL through this definition indicates the location with strongest suction of fluid element. Figure \ref{fig:bdcrossing} (a) shows the normalized momentum BL thickness versus rescaled aspect ratio $\Gamma/\Gamma_{opt}$ for different $Pr$. First, for $Pr \geq 0.5$, we observe that the momentum BL becomes thinner as $\Gamma$ is reduced towards $\Gamma_{opt}$. The amount of decline in thickness increases with $Pr$. Specially,  a $20\%$ decrease is seen for $Pr=0.5$ and a $50\%$ decrease for $Pr=40$. In contrast, such a decrease is absent for the cases of $Pr=0.1$ and $0.2$. We next examine the ratio of the thicknesses of the thermal and momentum BLs, $\lambda_T/\lambda_p$, versus $\Gamma/\Gamma_{opt}$, which is presented in Fig.~\ref{fig:bdcrossing} (b). It is found that the ratio of BLs thicknesses generally increases with decreasing $\Gamma$ for $Pr \geq 0.5$ but again, not for $Pr = 0.1$ and $0.2$. At a certain $\Gamma/\Gamma_{opt}$ around one, the thermal and momentum BLs have comparable thickness such that the location of the maximum normal stress coincides with that of maximum temperature fluctuation. This BLs crossing leads to the strongest coupling of normal stress and temperature fluctuation which is a favorable condition for plume emission  \citep{chong2017prl}.  However when the confinement becomes too severe, the momentum BL will be nested deeply inside the thermal one. Previous works have shown that temperature fluctuation decreases sharply within thermal BL \citep{belmonte1994pre,lui1998pre}. Therefore for momentum BL much thinner than the thermal BL, the suction of fluid will occur at positions lacking thermal instability and thus is not favorable to plume emission and heat transport. To provide evidence on such coupling, we have examined the normalized temperature standard deviation at the edge of momentum BL $\sigma_T/\sigma_{T,0}$ versus thickness ratio $\lambda_T/\lambda_p$ in figure \ref{fig:bdcrossingNu} (a). When the momentum BL becomes thinner and approaching the thickness of thermal BL, the increase of $\sigma_T$ at the edge of momentum BL  has been observed. However, when the thickness ratio becomes much larger than one, $\sigma_T$ drops sharply. As the figure suggests, BL crossing entails the strongest coupling between the normal stress and temperature standard deviation that is crucial to heat transport. To further demonstrate this, we plot $Nu$ against the thickness ratio in figure \ref{fig:bdcrossingNu} (b). It is seen that the optimal transport occurs for thickness ratio around one, namely between $1.08$ and $1.32$ (the shaded strip), and it again justifies that the BL crossing is intimately related to the optimal point.

\subsection{Global flow structures}
We now examine the changes in flow pattern brought by confinement for fluid with different $Pr$. Figure \ref{fig:meanfield} displays temporally-averaged mean velocity fields along x-z plane at mid-width together with temperature fields. We illustrate the properties of the mean flow field using four different $\Gamma$ 
($0.25$, $0.15$, $0.085$ and $0.025$) and three different $Pr\:$ ($0.1$, $1$ and $10$). We first consider the case of $Pr=0.1$ and $\Gamma=0.25$. It demonstrates a typical flow pattern in RB flow where there exists a well-defined large-scale circulation (LSC) with two counter-rotating corner rolls. For such a case, plumes most likely detach near sidewalls therefore hot and cold regions form on either side and the mean temperature field provides such a footmark. Due to the existence of corner rolls, plumes no longer impinge the opposite plates head on but rather at an angle as they are being steered. As $\Gamma$ decreases to $\Gamma=0.15$ and further to $0.085$, besides the reduced flow speed, the shape change of LSC is observed. As the figure suggests, the size of the LSC shrinks while the two corner rolls grow under confinement. With further confinement to $\Gamma=0.025$, the shape of LSC distorts largely and impingement of plumes becomes head on.

When $Pr$ is increased to $1$, the variation of flow pattern under confinement exhibits a different behavior. For $\Gamma=0.25$ at this $Pr$, the LSC still persists but larger corner rolls are seen as compared to the cases with $Pr=0.1$. However, when $\Gamma$ is further reduced to $0.15$, the mean field displays a four-roll pattern as opposed to a single-roll flow structure. Some earlier studies had  revealed that a time-averaged four-roll pattern can be brought about by the superposition of two flow fields with  opposite flow directions \citep{sun2005pre,kaczorowski2014jfm}. The change of flow pattern indicates that the LSC becomes less stable and therefore flow reversals occur more frequently. Under further confinement to $\Gamma=0.085$, the four-roll flow pattern still persists until $\Gamma$ reaches $0.025$ where the flow is dominated by vertical motion such that columnar structures appear in the temperature field. For $Pr=10$, single-roll structure has already broken down to four-roll at $\Gamma=0.25$. It demonstrates that the LSC becomes more unstable for larger $Pr$. With further confinement to $\Gamma=0.025$, it again exhibits the formation of columnar plumes and vertically aligned flow structures.

\begin{figure}[!h]
\includegraphics[width=0.8\textwidth]{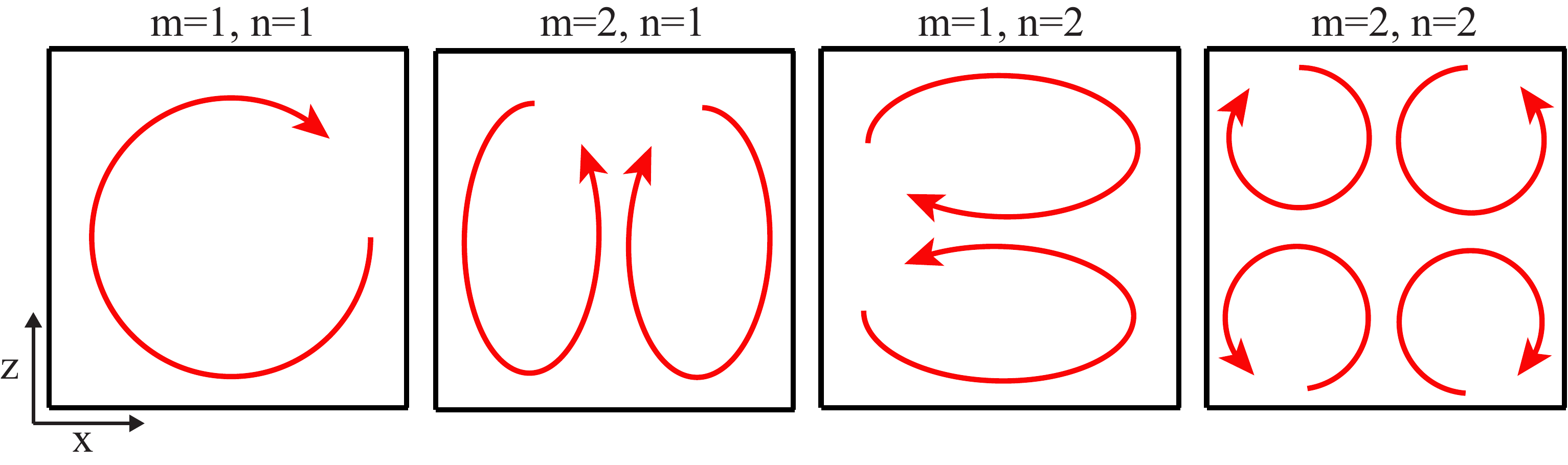}
\caption{\label{fig:podtypical}Schematic diagram of the four 2D Fourier modes ($u_x^{m,n}$,$u_z^{m,n}$) where $m, n$ equals to 1 or 2.}
\end{figure}

A more quantitative approach to study flow pattern and the strength of LSC is via the so-called 2D mode decomposition method \citep{Petschel2011,chandra2011pre,wagner2013pof}. We remark that this 2D technique is suitable here because our configurations are quasi-2D and the circulations are along the x-z plane. We apply the decomposition algorithm to a set of vertical cross-sections from instantaneous flow fields taken at mid-width, with the planar velocity field ($u_x$, $u_z$) being projections into the Fourier modes given as follows

\begin{equation}
\label{eq:ux}
u_x^{m,n} = 2\sin(m\pi x)\cos(n\pi z),
\end{equation}
\begin{equation}
\label{eq:uz}
u_z^{m,n} = -2\cos(m\pi x)\sin(n\pi z),
\end{equation}
where the first four modes are often considered \citep{chandra2011pre,wagner2013pof}, i.e. $m, n \in \{1, 2\}$. To give impression on those modes, we have drawn their flow pattern accordingly in figure \ref{fig:podtypical}. As suggested in \citep{wagner2013pof}, the projection is done component-wise on individual snapshots such that  the time series of $A_x^{m,n}(t)=\langle u_x(t) u_x^{m,n} \rangle_{x,z}$ and $A_z^{m,n}(t)=\langle u_z(t) u_z^{m,n} \rangle_{x,z}$ are obtained. Then a value $M^{m,n}$ representing the mode contribution can be evaluated by $M^{m,n} = \langle \sqrt{(A_x^{m,n})^2+(A_z^{m,n})^2} \rangle_t$.

\begin{figure}[!h]
\includegraphics[width=0.9\textwidth]{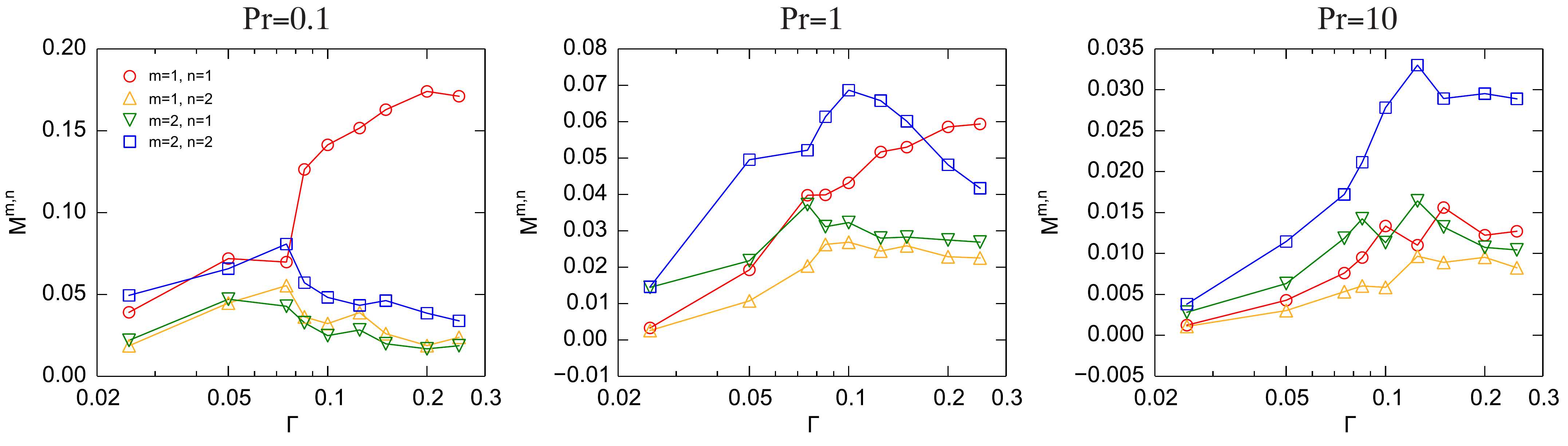}
\caption{\label{fig:pod}Magnitudes of the four 2D modes $M^{m,n}$ as a function of the aspect ratio $\Gamma$ for $Pr=0.1$, $1$ and $10$.}
\end{figure}

\begin{figure}[!h]
\includegraphics[width=0.6\textwidth]{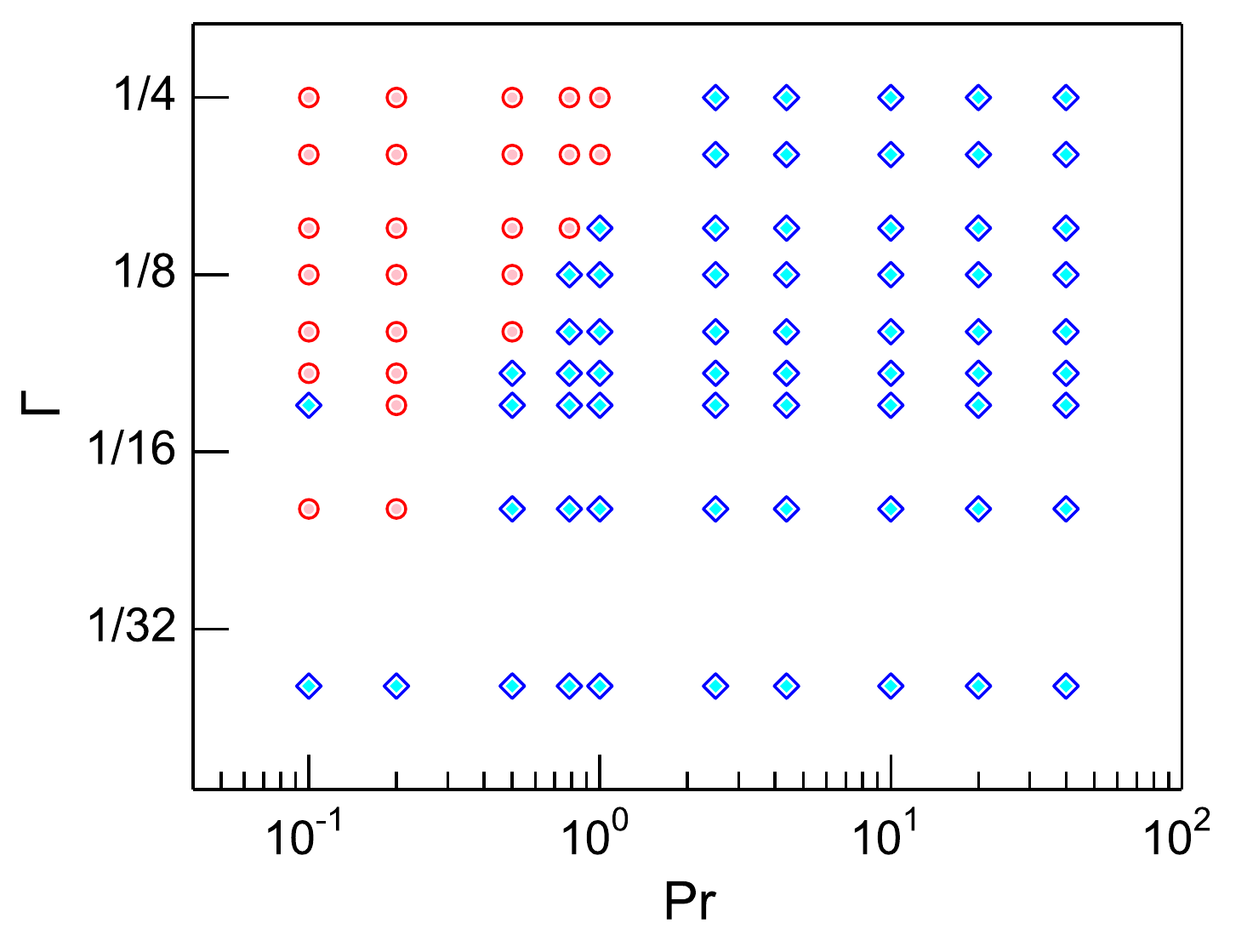}
\caption{\label{fig:phase}Phase diagram showing the existence of the LSC in the $\Gamma-Pr$ parameter space. Red circles represent cases where the value of  $M^{1,1}$ (single-roll mode) is the largest compared to the other three modes, otherwise, the cases are represented by blue diamonds.}
\end{figure}

In Fig. \ref{fig:pod} we compare the contributions of each mode $M^{m,n}$ (for $m, n \in \{1, 2\}$) as a function of $\Gamma$ for different $Pr$. Similar to the mean field, we have chosen $Pr=0.1$, $1$ and $10$ for demonstration. First we consider $Pr=0.1$ at $\Gamma=0.25$. The value of $M^{1,1}$, which represents the single-roll structure, is at least $4$ times larger than that of other modes. Upon decreasing $\Gamma$, the first mode $M^{1,1}$ becomes less dominant over other modes but is still the largest one. When $\Gamma$ reaches about $0.075$, $M^{1,1}$ declines sharply and becomes comparable to that of $M^{2,2}$, which represents the four-roll structure. From the discussion in the previous section, this is an indication that the LSC is being suppressed by confinement. When confinement is increased further to $\Gamma=0.025$, the magnitudes of $M^{1,1}$ and $M^{2,2}$ remain comparable. For $Pr=1$, $M^{1,1}$ is still the mode with largest magnitude at $\Gamma=0.25$ but its value is only $1.5$ times larger than that of others. The most prominent feature here is that the $M^{2,2}$ mode overtakes the $M^{1,1}$ mode  below $\Gamma=0.15$. Our quantitative result again demonstrates that the LSC is less stable for larger $Pr$. The mode $M^{2,2}$ dominates until $\Gamma=0.025$ where $M^{2,1}$ becomes comparable to $M^{2,2}$ which could indicate the formation of vertically aligned flow structures. For $Pr=10$, the mode $M^{2,2}$ is dominant over the full range of explored $\Gamma$ except $\Gamma=0.025$ at which magnitudes of $M^{2,1}$ and $M^{2,2}$ are comparable. This method enables us to quantify the strength of LSC and judge the presence of LSC by considering whether $M^{1,1}$ is the largest mode among others. Figure \ref{fig:phase} gives the phase diagram illustrating when to expect the existence of LSC in the $\Gamma-Pr$ parameter space. 

\section{Conclusions} \label{conclusions}
To summarize, we have studied the role played by the Prandtl number $Pr$ in geometrical confinement in terms of its effect on heat transport and flow structures, through DNS with $0.1 \leq Pr \leq 40$, $0.025 \leq \Gamma \leq 0.25$ and $Ra$ fixed at $10^8$. With regard to global convective heat transport, it is found that the existence, and the amount, of heat transport enhancement brought about by confinement depends strongly on $Pr$. For $Pr \geq 0.5$, significant heat transport enhancement has been observed but not for $Pr=0.1$ and $0.2$. We can define an optimal aspect ratio $\Gamma_{opt}$ at which $Nu$ is maximized for cases with enhancement. The $Pr$-dependence of $\Gamma_{opt}$ is found to follow a power-law relationship, $\Gamma_{opt}=0.11Pr^{-0.06}$. With regard to the amount of enhancement, the maximum enhancement generally increases with $Pr$ over the explored parameter range, ranging from $5.3\%$ to $15.3\%$ as $Pr$ increased from $0.5$ to $40$ (for $Ra = 10^8$; as the maximum enhancement also depends on $Ra$). Through the joint probability density function between vertical velocity and temperature standard deviations at mid-height, we can identify two competing effects due to confinement. On one hand, confinement reduces the flow strength;  on the other hand, plumes become more coherent as revealed by the temperature distribution function. However, besides these bulk properties, local quantities including plume coverage and temperature standard deviation at the edge of thermal BL seem to play more important roles in determining the global heat transport. It helps us to understand why larger $Pr$ produces larger enhancement and why no enhancement observed for smaller values of $Pr$, such as $0.1$ and $0.2$. To explain $\Gamma_{opt}$, we have examined the relative thickness of thermal and momentum BL $\lambda_T/\lambda_p$. It has been suggested in \citep{chong2017prl} that the BL thickness ratio closing to unity actually corresponds to the situation with strongest coupling between normal stress, i.e. suction of fluid, and the temperature fluctuations. This optimal coupling between the two quantities leads to the optimal transport. Our results with different $Pr$ at $Ra=10^8$ support this physical picture and again justify that optimal transport occurs when $\lambda_T/\lambda_p$ is around one.
We have in addition studied the global flow structure by examining the temporally averaged flow fields and through a 2D mode decomposition. A consequence of the geometrical confinement is the weakening of LSC, which is manifested by the fact that the single-roll flow structure is replaced by a four-roll structure in the time-averaged flow field. It is also supported by results from the 2D mode decomposition that $M^{1,1}$ mode no longer dominates. We further show that LSC persists over a wider range of $\Gamma$ for smaller $Pr$. A  phase diagram for the LSC existence is shown in figure \ref{fig:phase}.

This work was conducted under the Cooperation Agreement between the Max Planck Society and the Chinese University of Hong Kong. It was supported by the Hong Kong Research Grants Council under the Project No. CUHK404513 and CUHK1430115, and a NSFC$/$RGC Joint Research Grant N\_CUHK437$/$15; and through a Hong Kong PhD Fellowship. OS and SW thank German Research Foundation (DFG) for the support under the grants Sh~405/3 and Sh~405/4 -- Heisenberg Fellowship.
The authors are also grateful for the support of computational resources by the Leibniz-Rechenzentrum Munich under the project {\it pr84pu}.

\newpage

\setlength{\LTcapwidth}{0.9\linewidth}
\begin{longtable}{@{\extracolsep{\fill}}ccccccccc@{}}
  \caption{Simulation parameters and the global convective heat flux (all at $Ra=10^8$). Columns from left to right indicate $Pr$, $\Gamma$, the number of grid points in the three spatial directions $N_z \times N_x \times N_y$, the averaged grid spacing compared to the Kolmogorov length scale (or Batchelor length scale) $\Delta_z/\eta_k$ (or $\Delta_z/\eta_b$), the number of grid points in the thermal ($N_T$) and momentum ($N_u$) boundary layers, compared to the requirement \cite{shishkina2010njp}, the averaging time  $t_{avg}$ in free fall time units and the Nusselt number $Nu$.}\\
\hline\hline
     $Pr$    & $\Gamma$        & $N_z \times N_x \times N_y$   &   $\Delta_z/\eta_k$ &   $\Delta_z/\eta_b$ & $N_T$ & $N_u$ & $t_{avg}$ & $Nu$ \\[3pt]
\hline
\endfirsthead
\hline\hline
     $Pr$    & $\Gamma$        & $N_z \times N_x \times N_y$   &   $\Delta_z/\eta_k$ &   $\Delta_z/\eta_b$ & $N_T$ & $N_u$ & $t_{avg}$ & $Nu$ \\[3pt]
\hline
\endhead
\hline
\endfoot
0.1 &	0.025 & 	768x768x28	& 0.77	& 0.24  &40/2   & 20/1  &  200	& 13.36$\pm$0.01 \\
    &	0.050 &  	768x768x42	& 0.84	& 0.27  &30/3   & 15/1  &  184	& 18.62$\pm$0.06 \\
    &	0.075 & 	768x768x64	& 0.87	& 0.27  &27/3   & 13/1  &  184	& 20.83$\pm$0.10 \\
    &	0.085 & 	768x768x68	& 0.89	& 0.28  &25/3   & 12/1  &  156	& 22.80$\pm$0.12 \\
    &	0.100 & 	768x768x84	& 0.90	& 0.28  &24/3   & 12/2  &  144	& 23.72$\pm$0.14 \\
    &	0.125 & 	768x768x100	& 0.91	& 0.29  &23/3   & 11/2  &  178	& 24.81$\pm$0.19 \\
    &	0.150 & 	768x768x128	& 0.91	& 0.29  &23/3   & 11/2  &  178	& 25.11$\pm$0.21 \\
    &	0.200 & 	768x768x168	& 0.91	& 0.29  &22/3   & 11/2  &  158	& 25.37$\pm$0.26 \\
    & 	0.250 & 	768x768x200	& 0.92	& 0.29  &22/3   & 11/2  &  141	& 26.14$\pm$0.35 \\

0.2	& 0.025 	&  560x560x18	&   0.75	&   0.34    & 28/2  &  17/1  & 200	&  13.78$\pm$0.02 \\
  	& 0.050 	&  560x560x32	&   0.85	&   0.38    & 19/3  &  12/2  & 320	&  21.41$\pm$0.04 \\
  	& 0.075 	&  560x560x46	&   0.89	&   0.40    & 16/3  &  10/2  & 320	&  25.49$\pm$0.10 \\
  	& 0.085 	&  560x560x52	&   0.90	&   0.40    & 16/3  &  9/2  & 320	&  26.74$\pm$0.12 \\
  	& 0.100 	&  560x560x60	&   0.91	&   0.41    & 15/4  &  9/2  & 320	&  27.95$\pm$0.14 \\
  	& 0.125 	&  560x560x74	&   0.92	&   0.41    & 15/4  &  9/2  & 160	&  28.86$\pm$0.17 \\
  	& 0.150 	&  560x560x88	&   0.91	&   0.41    & 15/4  &  9/2  & 320	&  28.15$\pm$0.20 \\
  	& 0.200 	&  560x560x116	&   0.91	&   0.41    & 15/4  &  9/2  & 480	&  28.47$\pm$0.22 \\
  	& 0.250 	&  560x560x144	&   0.91	&   0.41    & 15/4  &  9/2  & 160	&  28.35$\pm$0.31 \\

0.5	& 0.025 	&  384x512x48	&   0.71	&   0.50 & 29/3 & 24/2  & 674	&  14.90$\pm$0.04 \\
  	& 0.050 	&  384x512x48	&   0.81	&   0.57 & 20/3 & 16/3  & 1349	&  23.92$\pm$0.05 \\
  	& 0.075 	&  384x512x74	&   0.86	&   0.61 & 16/4 & 13/3  & 750	&  30.57$\pm$0.06\\
  	& 0.085 	&  384x512x78	&   0.87	&   0.61 & 8/4 & 6/3  & 210	&  31.89$\pm$0.08 \\
  	& 0.100 	&  384x512x98	&   0.88	&   0.62 & 15/4 & 12/3  & 880	&  32.87$\pm$0.07 \\
  	& 0.125 	&  384x512x100	&   0.87	&   0.62 & 15/4 & 12/3  & 564	&  32.70$\pm$0.09 \\
  	& 0.150 	&  384x512x128	&   0.87	&   0.62 & 15/4 & 12/3  & 681	&  32.58$\pm$0.07 \\
  	& 0.200 	&  384x512x160	&   0.87	&   0.61 & 15/4 & 12/3  & 588	&  31.75$\pm$0.06\\
  	& 0.250 	&  384x512x200	&   0.86	&   0.61 & 16/4 & 13/3  & 473	&  31.22$\pm$0.05 \\

0.786	& 0.025 	&  384x512x48	&  0.57	&   0.50 & 29/3 & 27/2  & 672	&  15.00$\pm$0.03 \\
    	& 0.050 	&  384x512x48	&  0.65	&   0.58 & 19/3 & 18/3  & 1336	&  24.83$\pm$0.02 \\
    	& 0.075 	&  384x512x74	&  0.70	&   0.62 & 15/4 & 14/4  & 739	&  32.74$\pm$0.03 \\
    	& 0.085 	&  384x512x78	&  0.70	&   0.62 & 8/4 & 7/4  & 210	&  33.39$\pm$0.04 \\
    	& 0.100 	&  384x512x98	&  0.71	&   0.63 & 14/4 & 13/4  & 750	&  34.42$\pm$0.06 \\
    	& 0.125 	&  384x512x100	&  0.70	&   0.62 & 15/4 & 13/4  & 749	&  34.09$\pm$0.06 \\
    	& 0.150 	&  384x512x128	&  0.70	&   0.62 & 15/4 & 13/4  & 700	&  33.99$\pm$0.06 \\
    	& 0.200 	&  384x512x160	&  0.70	&   0.62 & 15/4 & 14/4  & 611	&  32.44$\pm$0.03 \\
    	& 0.250 	&  320x320x96	&  0.83	&   0.73 & 8/4 & 7/3  & 520	&  31.47$\pm$0.06 \\
1	& 0.025 	&  384x512x48	&   0.51	&   0.51	&  28/3  &  28/3  &   786&  15.42$\pm$0.03 \\
 	& 0.050 	&  384x512x48	&   0.58	&   0.58	&  19/3  &  19/3  &   934&  25.23$\pm$0.05\\
 	& 0.075 	&  384x512x74	&   0.62	&   0.62	&  15/4  &  15/4  &   996&  32.88$\pm$0.04 \\
 	& 0.085 	&  384x512x78	&   0.63	&   0.63	&  8/4  &  7/4  &   420&  34.41$\pm$0.08 \\
 	& 0.100 	&  384x512x98	&   0.63	&   0.63	&  14/4  &  14/4  &   972&  35.15$\pm$0.06 \\
 	& 0.125 	&  384x512x100	&   0.63	&   0.63	&  14/4  &  14/4  &   817&  34.83$\pm$0.04 \\
 	& 0.150 	&  384x512x128	&   0.62	&   0.62	&  15/4  &  14/4  &   748&  34.10$\pm$0.05 \\
 	& 0.200 	&  384x512x160	&   0.62	&   0.62	&  15/4  &  15/4  &   707&  32.90$\pm$0.05 \\
 	& 0.250 	&  384x512x200	&   0.62	&   0.62	&  15/4  &  15/4  &   837&  32.11$\pm$0.05 \\
 2.5& 0.025 	&  384x512x48	&   0.33	&   0.52	& 26/3  & 33/4  &1126  & 17.24$\pm$0.01\\
  	& 0.050 	&  384x512x48	&   0.37	&   0.59	& 18/3  & 23/5  &2000  & 27.42$\pm$0.04\\
  	& 0.075 	&  384x512x74	&   0.39	&   0.62	& 15/4  & 19/5  &989  & 33.79$\pm$0.03\\
  	& 0.085 	&  384x512x78	&   0.40	&   0.63	& 7/4  & 9/5  &420  & 35.51$\pm$0.11\\
  	& 0.100 	&  384x512x98	&   0.40	&   0.63	& 14/4  & 18/5  &1086  & 35.73$\pm$0.04\\
  	& 0.125 	&  384x512x100	&   0.40	&   0.63	& 14/4  & 18/5  &1009  & 35.44$\pm$0.03\\
  	& 0.150 	&  384x512x128	&   0.40	&   0.63	& 14/4  & 18/5  &829  & 34.83$\pm$0.04\\
  	& 0.200 	&  384x512x160	&   0.39	&   0.62	& 15/4  & 19/5  &669  & 33.50$\pm$0.04\\
  	& 0.250 	&  384x512x200	&   0.39	&   0.62	& 15/4  & 19/5  &538  & 32.92$\pm$0.04\\
4.38	&  0.025 	&  256x256x16	&   0.38	&   0.80  &11/3  &16/5   &300	&  18.35$\pm$0.01\\
    	&  0.050 	&  256x256x20	&   0.43	&   0.90  &7/4  &11/6   &400	&  28.70$\pm$0.08\\
    	&  0.075 	&  256x256x28	&   0.45	&   0.93  &9/4  &13/6   &500	&  33.51$\pm$0.08\\
    	&  0.085 	&  256x256x28	&   0.45	&   0.93  &6/4  &9/7   &300	&  35.14$\pm$0.11\\
    	&  0.100 	&  256x256x36	&   0.45	&   0.95  &6/4  &9/7   &640	&  35.41$\pm$0.11\\
    	&  0.125 	&  256x256x38	&   0.45	&   0.94  &6/4  &9/7   &370	&  34.63$\pm$0.18\\
    	&  0.150 	&  256x256x68	&   0.45	&   0.94  &9/4  &13/6   &350	&  34.13$\pm$0.07\\
    	&  0.200 	&  256x256x68	&   0.45	&   0.93  &6/4  &9/6   &350	&  33.30$\pm$0.08\\
    	&  0.250 	&  256x256x72	&   0.44	&   0.92  &7/4  &10/6   &380	&  32.20$\pm$0.09\\
 10	&  0.025 	&  256x256x8	&   0.17	&   0.53  & 20/3  & 36/6  & 413	&  18.41$\pm$0.11\\
  	&  0.050 	&  256x256x16	&   0.19	&   0.60  & 10/4  & 17/8  & 600	&  29.62$\pm$0.07\\
  	&  0.075 	&  256x256x24	&   0.20	&   0.63  & 9/4  & 16/9  & 510	&  34.26$\pm$0.08\\
  	&  0.085 	&  256x256x24	&   0.20	&   0.63  & 6/4  & 11/9  & 508	&  35.20$\pm$0.12\\
  	&  0.100 	&  256x256x26	&   0.20	&   0.63  & 8/4  & 14/9  & 375	&  34.98$\pm$0.10\\
  	&  0.125 	&  256x256x26	&   0.20	&   0.63  & 9/4  & 16/8  & 180	&  33.88$\pm$0.04\\
  	&  0.150 	&  256x256x42	&   0.20	&   0.62  & 9/4  & 16/8  & 345	&  32.85$\pm$0.06\\
  	&  0.200 	&  256x256x60	&   0.19	&   0.62  & 9/4  & 16/8  & 270	&  32.19$\pm$0.08\\
  	&  0.250 	&  256x256x64	&   0.19	&   0.61  & 9/4  & 16/8  & 560	&  31.65$\pm$0.10\\
 20	&  0.025 	&  256x256x8	&   0.18	&   0.81	&  12/3  & 27/8  &   490  &18.82$\pm$0.11\\
  	&  0.050 	&  256x256x16	&   0.20	&   0.91	&  10/4  & 21/10  &   945  &30.10$\pm$0.04\\
  	&  0.075 	&  256x256x24	&   0.21	&   0.95	&  9/4  & 18/11  &   751  &35.25$\pm$0.13\\
  	&  0.085 	&  256x256x24	&   0.21	&   0.95	&  6/4  & 13/11  &   509  &35.81$\pm$0.12\\
  	&  0.100 	&  256x256x26	&   0.21	&   0.95	&  7/4  & 16/11  &   375  &35.53$\pm$0.15\\
  	&  0.125 	&  256x256x34	&   0.21	&   0.93	&  9/4  & 19/11  &   687  &33.43$\pm$0.05\\
  	&  0.150 	&  256x256x42	&   0.21	&   0.92	&  9/4  & 20/10  &   728  &32.09$\pm$0.04\\
  	&  0.200 	&  256x256x60	&   0.21	&   0.92	&  9/4  & 20/10  &   378  &31.89$\pm$0.02\\
  	&  0.250 	&  256x256x64	&   0.21	&   0.92	&  9/4  & 19/10  &   300  &31.19$\pm$0.10\\
40	&  0.025 	&  256x256x8	&   0.13	&   0.81	&  12/3  &  32/10  &  875 &  19.40$\pm$0.14\\
 	&  0.050 	&  256x256x16	&   0.14	&   0.91	&  10/4  &  24/13  &  560 &  30.44$\pm$0.05\\
 	&  0.075 	&  256x256x24	&   0.15	&   0.95	&  9/4  &  22/14  &  952 &  35.75$\pm$0.05\\
 	&  0.085 	&  256x256x24	&   0.15	&   0.95	&  6/4  &  16/14  &  270 &  37.04$\pm$0.05\\
 	&  0.100 	&  256x256x26	&   0.15	&   0.95	&  7/4  &  20/14  &  1000 &  35.70$\pm$0.05\\
 	&  0.125 	&  256x256x34	&   0.15	&   0.94	&  9/4  &  23/13  &  700 &  33.71$\pm$0.02\\
 	&  0.150 	&  256x256x42	&   0.15	&   0.93	&  9/4  &  23/13  &  322 &  32.12$\pm$0.06\\
 	&  0.200 	&  256x256x60	&   0.15	&   0.92	&  9/4  &  24/13  &  504 &  31.37$\pm$0.04\\
 	&  0.250 	&  256x256x64	&   0.15	&   0.92	&  9/4  &  22/13  &  280 &  32.11$\pm$0.16\\
  \label{tab:sim}
\end{longtable}

\clearpage

\end{document}